\documentclass[%
 reprint,
 amsmath,amssymb,
 aps,
]{revtex4-2}

\usepackage{graphicx}
\usepackage{dcolumn}
\usepackage{bm}

\usepackage{hyperref}
\usepackage{nameref}
\newcommand{\RNum}[1]{\uppercase\expandafter{\romannumeral #1\relax}}
\usepackage[title]{appendix}

\usepackage{xcolor} 
\newcommand{\revision}[1]{#1}

\begin{document}

\preprint{APS/123-QED}

\title{Symmetry and stability of orientationally ordered collective motions of self-propelled, semiflexible filaments}

\def\jhuaffil{Department of Physics and Astronomy, Johns Hopkins University, Baltimore, Maryland 21218, USA}
\author{Madhuvanthi Guruprasad Athani}
\affiliation{%
 \jhuaffil
}%
\author{Daniel A. Beller}%
\email[]{d.a.beller@jhu.edu}
\affiliation{%
 \jhuaffil
}%


\begin{abstract}
Ordered, collective motions commonly arise spontaneously in systems of many interacting, active units, ranging from cellular tissues and bacterial colonies to self-propelled colloids and animal flocks. Active phases are especially rich when the active units are sufficiently anisotropic to produce liquid crystalline order and thus active nematic phenomena, with important biophysical examples provided by cytoskeletal filaments including microtubules and actin. Gliding assay experiments have provided a testbed to study the collective motions of these cytoskeletal filaments and unlocked diverse collective active phases, including states with long-range orientational order. However, it is not well-understood how such long-range order emerges from the interplay of passive and active aligning mechanisms. We use Brownian dynamics simulations to study the collective motions of semiflexible filaments that self-propel in quasi-two dimensions, in order to gain insights into the aligning mechanisms at work in these gliding assay systems. We find that, without aligning torques in the microscopic model, long-range orientational order can only be achieved when the filaments are able to overlap. The symmetry (nematic or polar) of the long-range order that first emerges is shown to depend on the energy cost of filament overlap and on filament flexibility. However, our model also predicts that a long-range-ordered active nematic state is merely transient, whereas long-range polar order is the only active dynamical steady state in systems with finite filament rigidity.
\end{abstract}

\maketitle


\section{\label{sec:level1} Introduction}

Active matter systems are composed of a large number of interacting ``agents'' that individually convert internal energy into motion. Flocks of birds, herds of animals, and growing or swarming bacterial colonies are some examples of active matter across various length scales \cite{VICSEK201271, Cavagna_starling_flocks, animal_collective_behavior, collective_motion_humans}. The formation of spontaneously ordered collective motion is a common phenomenon exhibited by active systems. Clustering and flocking ordered states seen in many active systems can be modeled using active Brownian particles, in cases where the active agents are spherically symmetric \cite{Vicsek, Cates2015, Active_Brownian_Disks_MIPS2018}. But active matter also comprises anisotropic systems with agents whose aspect ratios differ strongly from unity, such as swimming rod-shaped \textit{bacillus subtilis} \cite{Zhang_bacterial_swarms} or \textit{E.~coli} bacterial swarms \cite{E.coli_swarm1994, Bacterial_swarming_review2009}, \revision{glioma cells in brain tumors \cite{Jamous2020, Comba2022},} and cytoskeletal filaments such as actin and microtubules \cite{Sanchez2012, Schaller2010, actin_Molloy_2010}. The anisotropy of these individual active particles is responsible for the orientational order and the distinct behaviors of active nematics \cite{Doostmohammadi_Active_nematics2018, Marchetti_hydrodynamics_review}. 


Collective phases of anisotropic cytoskeletal biofilaments have been experimentally observed in gliding assay systems, which were originally used to study molecular motors such as kinesin, dynein, and myosin, which walk along cytoskeletal filaments such as actin and microtubules \cite{first_gliding_assay_Allen1985, Hunt1994, Phillips2012}. In a conventional \textit{in vitro} gliding assay setup, molecular motors are uniformly dispersed on a glass slide where they are held fixed with their tail end, which attaches to molecular cargo \textit{in vivo}, here attached to the surface of the glass slide. Cytoskeletal filaments glide on these motors, which push the filaments forward in a defined direction, making the system microscopically polar. When the densities of motors and filaments are sufficiently high in a gliding assay, diverse active collective phases spontaneously emerge, including polar flocks, nematic and polar lanes, asters, and ``spools" of filaments  \cite{depletion_gliding_assay_Inoue2015, Saito_globalOO, Sumino2012, Huber}. One such active state observed is the active nematic state with long-range orientational order. Understanding how this long-range ordered (LRO) nematic state emerges and evolves from microscopically polar, flexible constituents is the main subject of this paper.

The effects of shape on interacting colloidal particles were first studied in thermal systems of passive, hard, rod-like colloidal particle systems to reveal that higher aspect ratio and higher packing fraction favor long-range nematic order \cite{Onsager1949}. These trends hold also in active systems of self-propelled hard rods; additionally, increasing the self-propulsion strength favors nematic alignment \cite{Baskaran2008, Ginelli2010, Wensink2012, abkener2013}.

In gliding assays, however, the hard-rod approximation becomes unsuitable due to large deformations of the active filaments.  The effects of polymer flexibility on nematic ordering are notable already in passive, lyotropic liquid crystals, where lower rigidity disfavors order and increases the isotropic-nematic phase transition concentration, compared to hard-rod colloids \cite{KHOKHLOV1981546}. Rigidity is likewise an important parameter in determining the active phases of self-propelled, semiflexible filaments \cite{kim2018, Prathyusha, Huber, Betterton, Tanida, Zhou2022}.

Past simulation studies of active semiflexible filaments that have reported LRO states \cite{kim2018, Huber, Tanida} employed an attractive or aligning interaction between nearby active filaments in their microscopic model. These assumptions were justified by the use of depletants, which cause filaments to bundle in parallel or antiparallel orientations, in conventional gliding assay experiments where LRO states have been observed  \cite{Tanida, Huber, depletion_gliding_assay_Inoue2015, Saito_globalOO}. However, recent experiments with microtubules on lipid substrates have reported LRO states without the need for depletants \cite{Memarian}. This experimental evidence motivates us to study how LRO active nematic states emerge and evolve without imposed aligning or attractive interactions in the microscopic model. 

Another important feature of experimental gliding assays is that filaments frequently cross over one another, with one filament partially and temporarily delaminating from the substrate \cite{depletion_gliding_assay_Inoue2015, Huber, Memarian, Zhou2022, Tanida}. These crossovers are forbidden by models that assume strictly two-dimensional motion of impenetrable filaments. As we show in this work, filament crossovers are essential to LRO states in the absence of aligning or attractive interactions at the filament-pair level.

In this paper, we conduct a detailed study of the active phase behaviors of self-propelled, semiflexible filaments without imposed attractive or aligning interactions between the active agents. We determine the conditions under which these filaments develop collective motions with long-range orientational order, by varying parameters including the filaments' flexibility and area fraction.

We find that a long-range orientationally ordered phase does not arise when the filaments are strictly constrained to two dimensions through volume exclusion. Instead, we observe LRO states only in \textit{quasi}-two dimensional systems, in which we replace the volume-excluding steric repulsion with a finite energy cost for filament overlap \cite{Betterton,Ghosh2022}, in order to account for crossovers. 

In our simulations, the symmetry of the long-range orientational order that first emerges may be polar or nematic (apolar) depending on the flexibility of the filaments. Surprisingly, we find the LRO active nematic state to be merely transient, with all LRO systems eventually becoming polar. Our model predicts that the time taken by the system to reach the LRO active polar steady state monotonically increases with filament stiffness. These findings suggest that experimental observations of LRO active nematic states in microtubule gliding assays may be better interpreted as transient behavior, observable due to the finite lifetime of experiments and relatively high rigidity of the filaments.

\section{\label{sec:level2} Model}
We use Brownian dynamics to simulate active (self-propelled), semiflexible filaments in two dimensions. Each filament is modeled as a bead-spring chain with a Hookean restoring force between adjacent beads and a bending stiffness $\tilde{\kappa}$. Each bead self-propels with a constant active force in the chain's local tangent direction, resulting in an active \revision{speed} 
$v_0$ in the absence of other forces. The local tangent direction is calculated at each time step from bead positions and does not have its own relaxation time in this model.

Steric repulsion between beads of different chains is modeled using a modified Weeks-Chandler-Andersen potential \cite{Prathyusha, WCA1}:

\begin{multline}
    U_{\mathrm{WCA}}\left(r_{ij}\right)   \\ 
    = \begin{cases}
            4 \varepsilon\left[
                \left(
                    \frac{\sigma}{r_{ij}+r_{\mathrm{shift}}}
                \right)^{12} - \left(
                    \frac{\sigma}{r_{ij}+r_{\mathrm{shift}}}
                \right)^{6}
            \right]+\varepsilon, 
            & 
            r_{ij} \leq 2^{1 / 6} \sigma 
            \\ 0, 
            & r_{ij}>2^{1 / 6} \sigma
    \end{cases} 
    \label{wca_eq}
\end{multline}

Here, $r_{ij}$ is the distance between the centers of two beads \revision{$i$ and $j$}, and $r_{\mathrm{shift}}$ is a parameter introduced to make the $U_{\mathrm{WCA}}$ potential finite when $r_{ij} = 0$; this ensures that chains can interpenetrate. The overlap of chains in our explicitly two-dimensional model mimics the crossover events of a quasi-two-dimensional microtubule gliding assay experimental system. In addition to acting between beads of different chains, $U_{\mathrm{WCA}}$ also acts between bead pairs on the same chain separated by more than three bead positions, penalizing self-intersections of the semiflexible chains. 

We scale simulation lengths in units of $\tilde x = 2^{1/6}\sigma$, the range of the WCA repulsive potential. In these units, the bead radius is $r_0 = 0.5 \tilde x$. The chain length $\ell = 31 r_0$ is the same for all chains and is kept fixed across all simulation results presented here; polydispersity will be investigated in future work.  Time $t$ is scaled with respect to $\tilde x/v_0$, although in our plots we present time \revision{$\tilde t$} in units of  $\ell / v_0$. The  Brownian (over-damped) dynamics updates bead positions through
\begin{equation}
  \textbf{x}(t + dt) = \textbf{x}(t) + \textbf{v}dt +  \sqrt{2Ddt} \boldsymbol{\xi}(t)
  \label{position} \revision{,}
\end{equation}
\begin{equation}
   \textbf{v} =   \textbf{F}_{\mathrm{total}}/\gamma \revision{.}
\end{equation}

The third term on the right-hand side of Eq.~\ref{position} is the stochastic noise term representing random fluctuations in bead positions, with each vector component drawn from a zero-mean Gaussian white noise distribution. We use a second-order stochastic Runge-Kutta (SRK-2) method \cite{BD_SRK}  to update the bead positions, with an adaptive time step. All simulations are conducted in a square box of side length $L_x$ with periodic boundary conditions in both the $x$ and $y$ directions. Full details of the bead-spring-chain model can be found in Appendix \ref{Appendix A}.

\section{Results}
\subsection{\label{sec:level3}Dependence of active state on chain penetrability}
\begin{figure*}
    \centering
    \includegraphics[width=17.5cm]{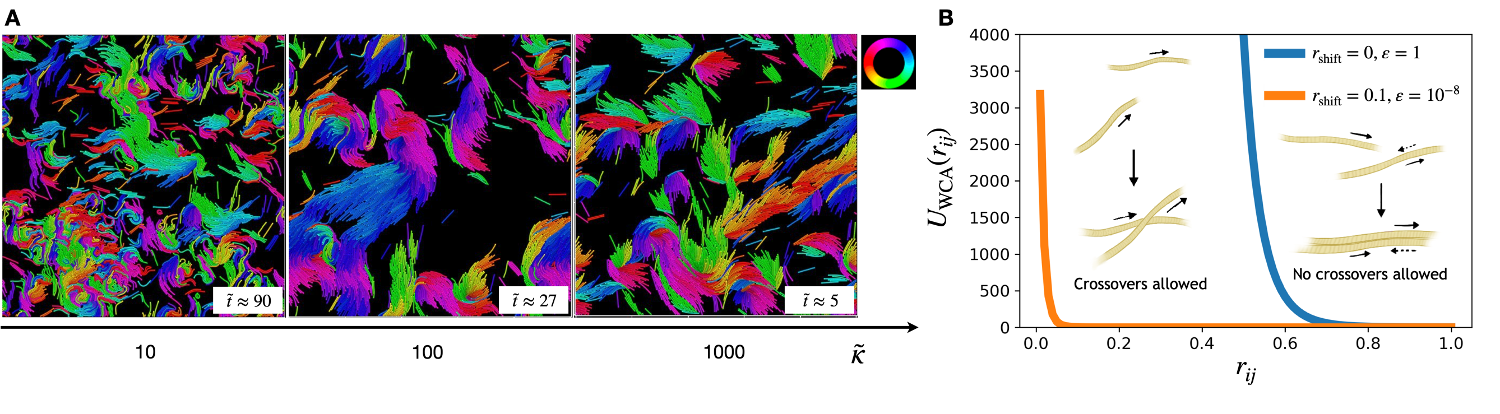}
    \caption{\textbf{\revision{A.}} Polar flocks observed when $r_{\mathrm{shift}} = 0$ and $\varepsilon = 1$ in $U_{\mathrm{WCA}}$. The simulation box size $L_x= 250$, area fraction $\phi = 0.5$, and number of chains $N=2016$. The color wheel legend, here and in subsequent figures, indicates the \revision{the orientation of each filament, averaged over the tangents at each bead}. \textbf{\revision{B.}} Plot of the WCA potential for bead interactions in the cases of steric repulsion (blue) and penetrable filaments (orange).} 
    \label{fig:polar_flocks}
\end{figure*}
We begin by examining the strictly two-dimensional scenario in which crossovers are prohibited, with $r_{\mathrm{shift}} = 0$. This choice makes the repulsion potential $U_{\mathrm{WCA}}(r_{ij})$ diverge when two chains overlap ($r_{ij} = 0$), thus forbidding chains to interpenetrate when they collide.  We vary bending rigidity $\tilde \kappa$, which in passive systems would favor nematic order as it is increased \cite{KHOKHLOV1981546}. However, in this active system, we find by varying $\tilde \kappa$ across three orders of magnitude that the system does not exhibit long-range order. Instead, polar flocks are seen for all values of $\tilde{\kappa}$, as shown in Fig.~\ref{fig:polar_flocks}A and in Supplemental Video 1 \cite{SI}.

These results indicate that long-range orientational order for mutually impenetrable active semiflexible chains cannot be achieved without an imposed alignment or attractive potential. Active polar flocks similar to those observed here have been reported in various microtubule and actin filament gliding assays both in experiments using depletants and in simulations where the filaments were strongly confined to two dimensions \cite{Tanida, Schaller2010, Weitz2015, abkener2013, Betterton, Gromann2020}. Similar polar flocks have also been reported in experiments on rod-shaped \textit{E.~coli} bacterial colonies when they were strongly confined to two dimensions \cite{Ghosh2022}.

We now introduce chain penetrability by modifying the $r_{ij}$-dependence of the $U_{\mathrm{WCA}}$ interaction potential as shown in Fig.~\ref{fig:polar_flocks}B, with $r_{\mathrm{shift}}$ fixed at $0.1$, so that the potential is finite at $r_{ij}=0$. We vary $\varepsilon$ to change the potential barrier for chain interpenetration, which affects the frequency of crossovers (Supplemental Video 2 \cite{SI}). 

\begin{figure*}
    \centering
    \includegraphics[width=17.5cm]{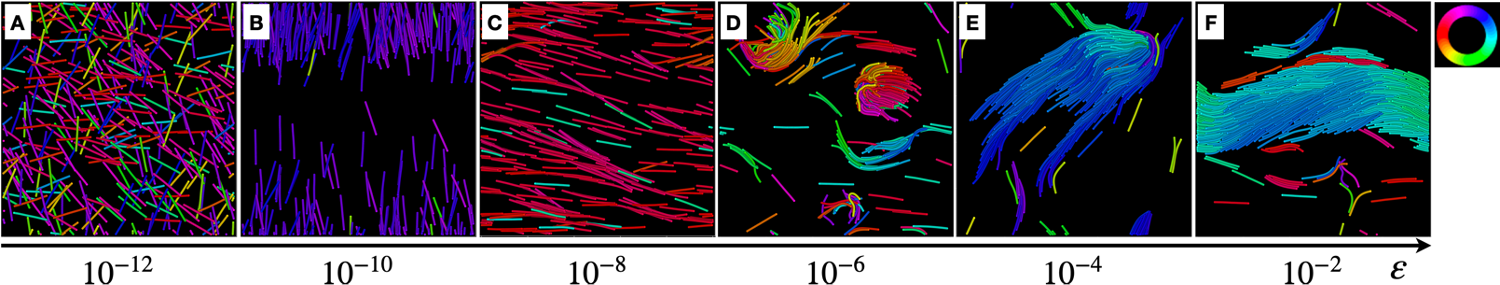}
    \caption{Changing the energy cost $\varepsilon$ for chains to \revision{interpenetrate} 
    we obtain the following states: \textbf{A.} Isotropic state with no positional or orientational order. \textbf{B.} Polar wave state where chains move in the same direction as the wave. \textbf{C.} Long-range orientational ordered state. The orange line in Fig. \ref{fig:polar_flocks}B corresponds to the potential used here. \textbf{D.} Jammed flock state. \textbf{E}-\textbf{F.} Dynamic polar flocks and lanes. All the simulations have box size $L_x= 100$, $\phi = 0.5$, $\ell = 15.5$, $\tilde{\kappa} = 30$ and $r_{\mathrm{shift}} = 0.1$. The color wheel legend indicates the orientation of each filament, averaged over the tangents at each bead.}
    \label{fig:wca_fig}
\end{figure*}

At the lowest $\varepsilon$ values tested, such as $\varepsilon = 10^{-12}$ as shown in Fig.~\ref{fig:wca_fig}\textbf{A}, we obtain an isotropic state, as chains simply cross over each other essentially unimpeded. When $\varepsilon = 10^{-10}$ we observe polar density waves (Fig.~\ref{fig:wca_fig}\textbf{B}) traveling parallel to the chain self-propulsion direction, as previously seen in simulations and experiments involving actin filaments \cite{Huber} (where the term ``polar clusters'' is applied to this state). A state with no positional order but long-range orientational order (Fig.~\ref{fig:wca_fig}\textbf{C}), our main interest in this work, is obtained when $\varepsilon = 10^{-8}$. We therefore maintain $U_{\mathrm{WCA}}$ parameters of $\varepsilon = 10^{-8}, r_{\mathrm{shift}} = 0.1$ in the following sections of this paper.

At even higher values of $\varepsilon$, we obtain density inhomogeneities in the form of flocks. When $\varepsilon = 10^{-6}$, a jammed flocking state appears (Fig.~\ref{fig:wca_fig}\textbf{D}), with characteristics distinct from the polar flocks seen in Fig.~\ref{fig:polar_flocks}\textbf{A}.  The jammed flocks seen here are formed by counter-moving chains that are dynamically arrested relative to the center of mass of the flock, which then moves as a whole. (In contrast, all the chains in a particular polar flock seen in Fig.~\ref{fig:polar_flocks}\textbf{A} are oriented nearly in the forward direction.) The jammed flocking state of  (Fig.~\ref{fig:wca_fig}\textbf{D}) can be interpreted as a form of motility-induced phase separation \cite{Cates2015}, with counter-moving filaments impeding each others' motions as they overlap.

The density inhomogeneities seen in Fig.~\ref{fig:wca_fig}\textbf{E} and \textbf{F} are dynamic and they reconfigure into different lanes and flocks multiple times in the course of the simulation. While the polar lane in Fig.~\ref{fig:wca_fig}\textbf{F} spans the system, we observe upon increasing the system size a state like the finite flock of Fig.~\ref{fig:wca_fig}\textbf{E}. We therefore interpret the system-spanning lane of Fig.~\ref{fig:wca_fig}\textbf{F} as a finite-size effect. 

\subsection{\label{sec:level4}Long-range-ordered states}
\begin{figure*}
    \centering
    \includegraphics[width=17.5cm]{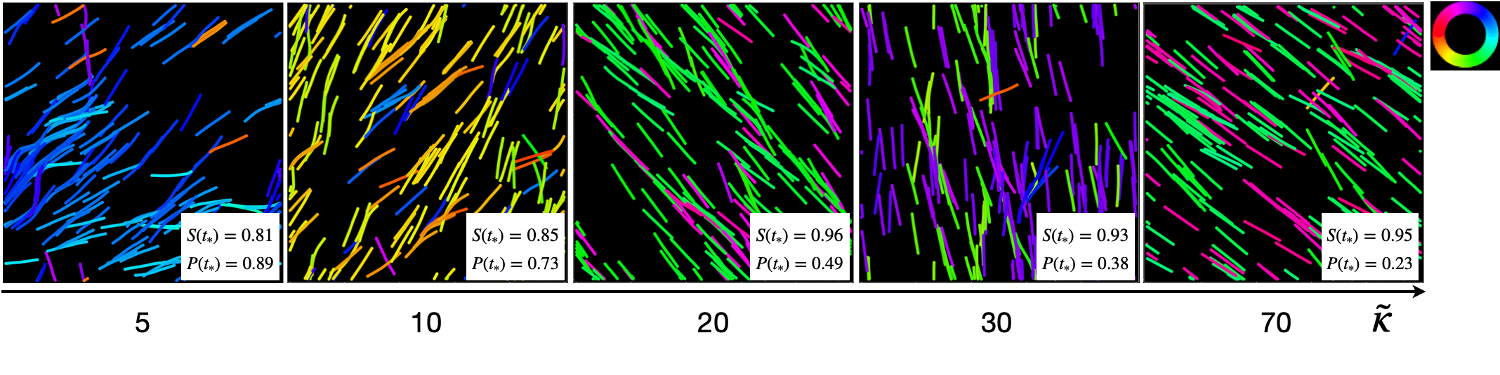}
        \caption{States with long-range orientational order obtained for different bending rigidities $\tilde{\kappa}$. All simulations have $L_x= 100$, $\phi = 0.3$, $\ell = 15.5$,  $\varepsilon = 10^{-8}$ and $r_{\mathrm{shift}} = 0.1$. $S(t_*)$ and $P(t_*)$ are the nematic and polar order parameters\revision{,} respectively\revision{,} of the simulation runs pictured here, at time $t_*$ roughly corresponding to the saturation time for nematic order. 
    }
    \label{fig:polar_nematic}
\end{figure*}

With $\varepsilon$ fixed at $10^{-8}$ henceforth, we now investigate active states with long-range orientational order (LRO) in greater detail. First, we study the effect of changing bending rigidity $\tilde \kappa$ on the form of the orientational order. In Fig.~\ref{fig:polar_nematic}, we see snapshots of the initially formed long-range orientational order (time $t_* = 200.58 [\ell/v_0]$) for a range of $\tilde{\kappa}$ values. To quantify the orientational order we compute two different order parameters: nematic order \revision{$S(\tilde t)=\left|\left\langle\exp \left(i 2 \theta_{j}^{\tilde t}\right)\right\rangle_{j}\right|$} and polar order \revision{$P(\tilde t)=\left|\left\langle\exp \left(i\theta_{j}^{\tilde t}\right)\right\rangle_{j}\right|$}, where $\theta$ is the average orientation of all the beads in a chain with respect to the positive $x$-axis, and subscripts $j$ indicate a system average over all chains. We label a state as orientationally ordered when \revision{$S(\tilde t) \geq 0.7$}. An orientationally ordered state is labeled a LRO polar state if \revision{$P(\tilde t) \geq 0.7$}, and a LRO nematic state otherwise.

\begin{figure*}
    {\includegraphics[width=16.5 cm]{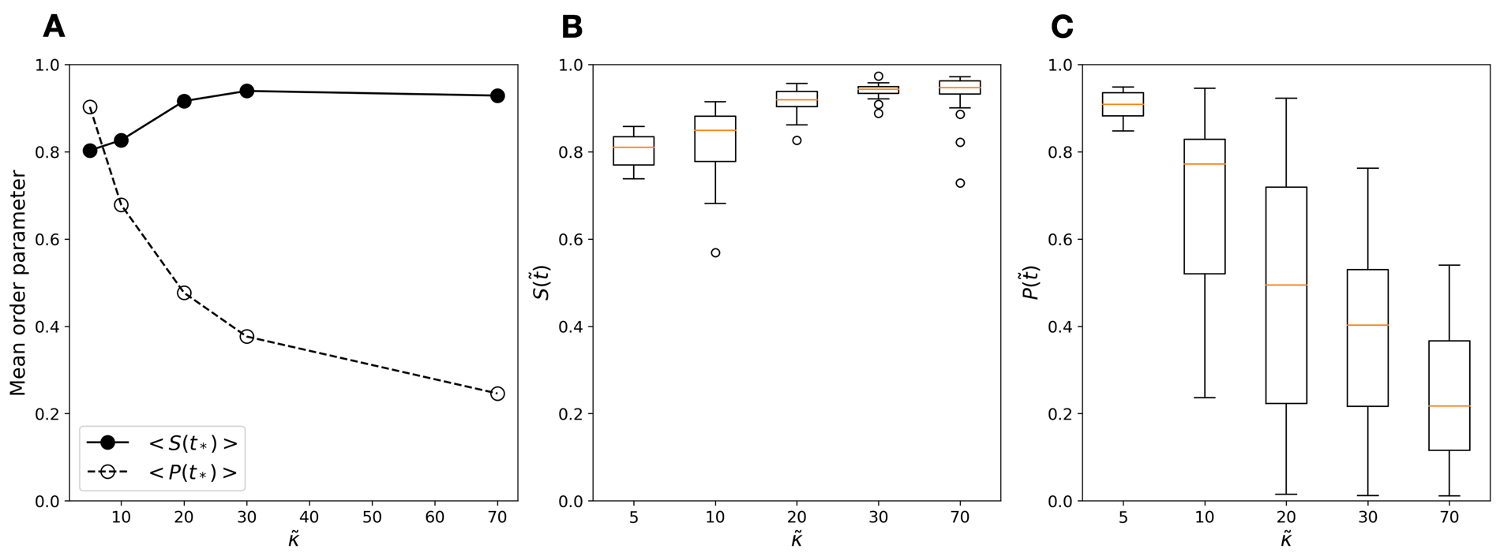}}  
    {\caption{\textbf{A.} Mean nematic order parameter and polar order parameter plotted against bending rigidity at time $t_*$ when all the simulations have reached long-range orientational order.
    \textbf{\revision{B.}} and \textbf{\revision{C.}} Median and the spread of the nematic (\textbf{B}) and polar (\textbf{C}) order values in 20-run ensembles at time $t_*$. }\label{fig:op}}
\end{figure*}

In Fig.~\ref{fig:polar_nematic}, we see that more rigid chains (greater $\tilde{\kappa}$) form a LRO nematic state ($S(t_*) > 0.7$ and $P(t_*) < 0.7$) whereas more flexible chains achieve a LRO polar state ($P(t_*) > 0.7$). 
The same trend is revealed when nematic and polar order parameters, averaged over 20-run ensembles, are plotted against $\tilde \kappa$  (Fig.~\ref{fig:op}\textbf{A}): whereas $\left<S(t_*)\right>$ increases modestly with increasing $\tilde \kappa$, $\left<P(t_*)\right>$ decreases dramatically. Biopolymers such as microtubules have a higher bending rigidity (persistence length $\approx$ 1 mm) compared to actin filaments (persistence length $\approx$ 10-20~$\mu$m) \cite{actin_mt_persistence}. While LRO polar states have not been observed in experiment, our results are consistent with the known tendency of actin gliding assays to exhibit short-range polar order whereas nematic order (short- or long-range) is more common in microtubules \cite{depletion_gliding_assay_Inoue2015, Memarian, Tanida, Huber, Schaller2010, Weitz2015}.

The modest increase in nematic order with increasing rigidity can be attributed to the fact that more rigid chains have greater restoring forces against bending deformations, and are therefore less likely to be significantly deflected from their direction of motion due to collisions or random fluctuations. Box plots in Fig.~\ref{fig:op}\textbf{B} and \textbf{C} show the median and spread of nematic and polar order parameter values at time $t_*$, when all the simulations have reached long-range orientational order. It is interesting to note that, at larger $\tilde \kappa$ values that gave apparently nematic states, the polar order parameter values (Fig.~\ref{fig:op}\textbf{C}) are significantly above zero and widely spread. This indicates that LRO nematic order coexists with LRO polar order, and raises the question of whether the LRO nematic state is, in fact, the dynamical steady state.\\

\subsection{\label{sec:level5}Nature of the active steady state}
\begin{figure*}
    \centering
    \includegraphics[width=15.5 cm]{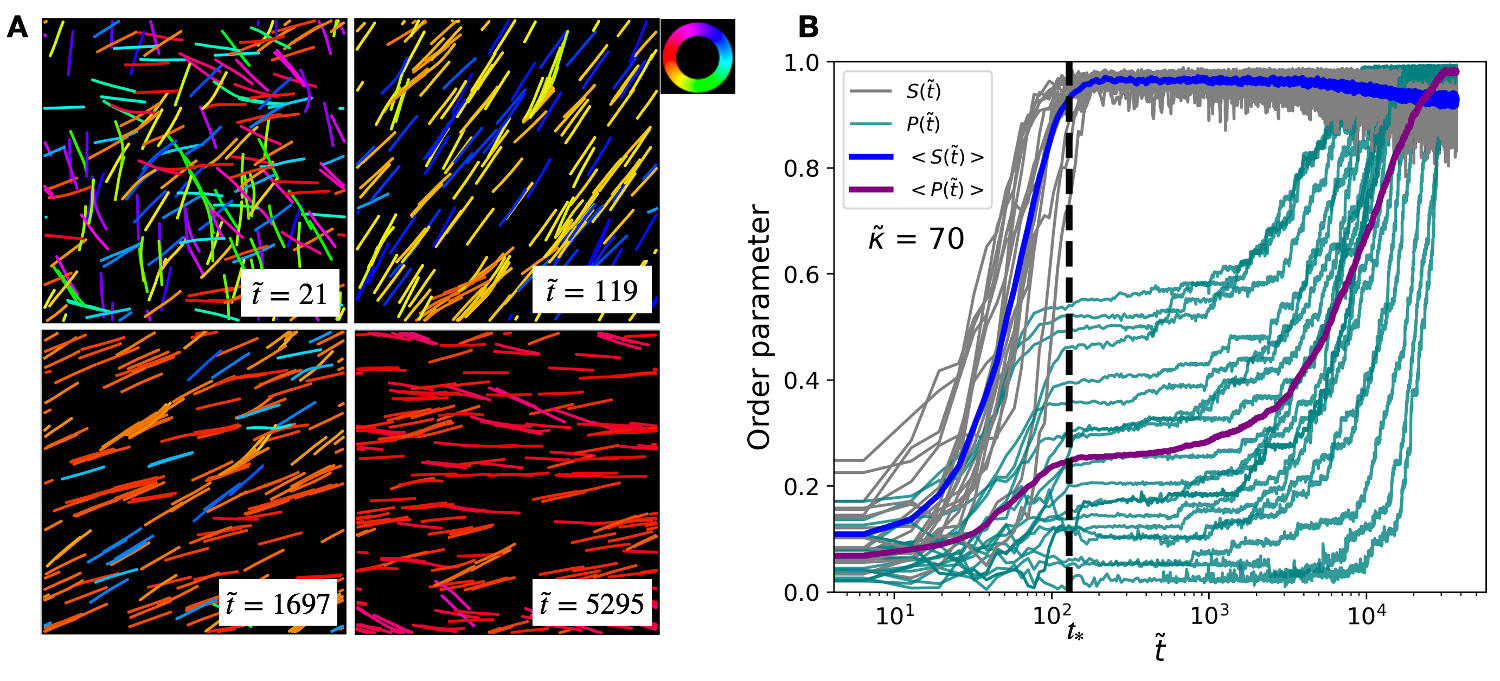}
    \caption{\textbf{\revision{A.}} Snapshots of the time-evolution of a system with $\tilde{\kappa} = 30$ and $\phi = 0.3$. At long times the order is polar, not nematic. \textbf{\revision{B.}} Polar (cyan) and nematic (grey) order parameters for an ensemble of simulations (n = 20) with $\tilde{\kappa} = 70$. Magenta and blue curves represent the ensemble-mean polar and nematic order parameters, respectively, at each time. All systems reach long-range orientational order around the same time $t_*$, marked by the black dashed line.}
    \label{fig:steady_state}
\end{figure*}

In order to determine whether LRO nematic order with weak LRO polar order is a dynamically stable active state, we now examine the evolution of orientationally ordered systems over times much longer than $t_*$. Examining the time-evolution represented by the snapshots in Fig.~\ref{fig:steady_state}\textbf{A} and Supplemental Video 3 \cite{SI}, we can see that the LRO active nematic state discussed in the previous section is merely transient, and the system reaches an active global polar steady state at longer times. To quantify this evolution of orientational order, we plot in Fig.~\ref{fig:steady_state}\textbf{B} the polar and nematic order parameters against time for each simulation in a 20-run ensemble, with bending rigidity $\tilde{\kappa} = 70$. The black dashed line indicates the time $t_*$ when all the simulations have achieved long-range nematic orientational order. However, the nematic (apolar) state is transient, as indicated by \revision{$P(\tilde t)$} rising to saturation near $1$ at much later times. The active steady state is therefore a LRO active polar state.  At intermediate times after $t_*$, the polar order appears to plateau for a significant time interval, though the plateau value of \revision{$P(\tilde t)$} exhibits wide variation between simulation runs.

\begin{figure*}
    \centering
    \includegraphics[width=17.5cm]{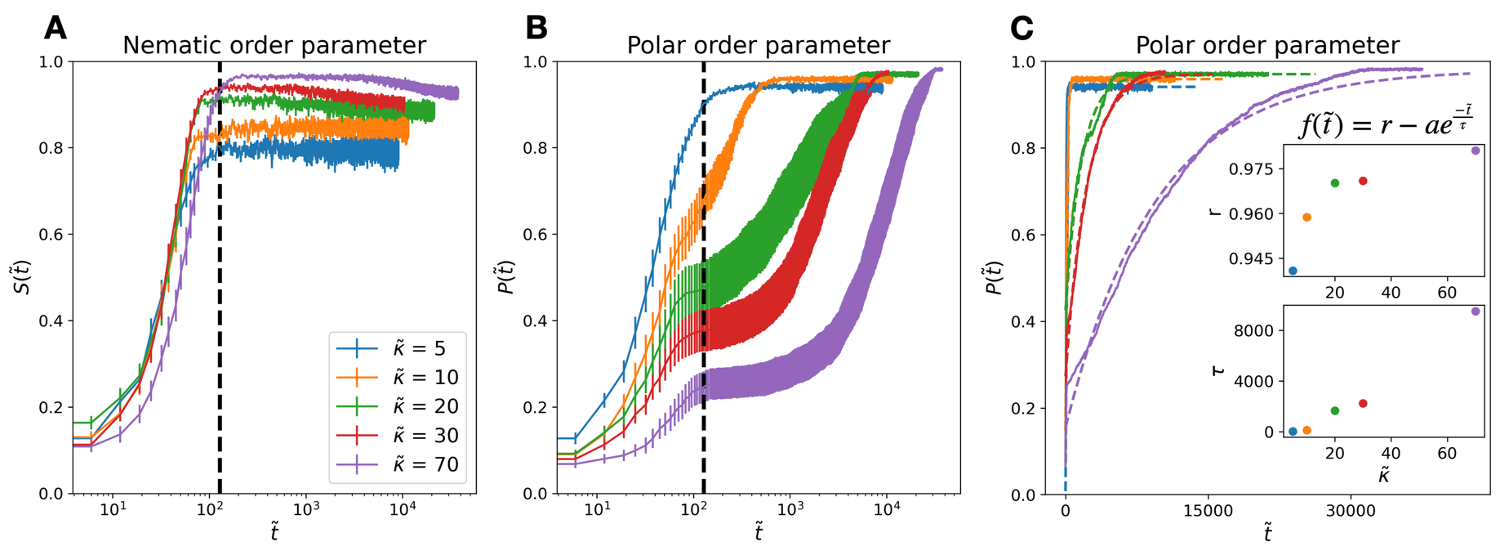}
    \caption{Ensemble-averaged ($n=20$) nematic order parameter and polar order parameter \revision{(area fraction $\phi = 0.3$)}. Error bars (appearing as widths of curves at larger time values) indicate the standard error of the mean. \textbf{A.} Global nematic order parameter \revision{$S(\tilde t)$} saturates at approximately the same time $t_*$ for all $\tilde{\kappa}$, values indicating long-range orientational order. \textbf{B.} Global polar order parameter \revision{$P(\tilde t)$} saturates over time for all $\tilde{\kappa}$, but at later times for larger $\tilde \kappa$. \textbf{C.} A saturating exponential function \revision{$f(\tilde t) = r-ae^{{-\tilde t}/{\tau}}$} (dashed curves) is fit to the ensemble-averaged \revision{$P(\tilde t)$} data (solid curves). Insets show \revision{fit parameters} $r$ (top) and $\tau$ (bottom) increasing with increasing $\tilde{\kappa}$.}
    \label{fig:steady_state_plots}
\end{figure*}

To understand the effect of changing bending rigidity on the steady-state long-range order, we plot in Fig.~\ref{fig:steady_state_plots} the ensemble-averaged nematic (Fig.~\ref{fig:steady_state_plots}\textbf{A}) and polar (Fig.~\ref{fig:steady_state_plots}\textbf{B}, \textbf{C}) order parameters against time for multiple values of $\tilde \kappa$. From Fig.~\ref{fig:steady_state_plots}\textbf{A} we see that nematic order saturates at around the same time $t_*$ for all $\tilde{\kappa}$ values. Fig.~\ref{fig:steady_state_plots}\textbf{B} shows that, while polar order eventually saturates for all $\tilde{\kappa}$ values, the time of saturation depends strongly on $\tilde{\kappa}$. Systems with larger $\tilde{\kappa}$ tend to take much longer times to reach their LRO active polar steady state. To examine the $\tilde \kappa$-dependence of the \revision{$P(\tilde t)$} saturation time, we plot in Fig.~\ref{fig:steady_state_plots}\textbf{C} the same data as in Fig.~\ref{fig:steady_state_plots}\textbf{B} but with a linear scale on the time axis. Fitting a saturating exponential function \revision{$f(\tilde t) = r-ae^{{-\tilde t}/{\tau}}$} to the \revision{ensemble-averaged} polar order parameter yields typical time scales $\tau$ for saturation of polar order at \revision{$P(\tilde t) \rightarrow r$}. As shown in the inset to Fig.~\ref{fig:steady_state_plots}\textbf{C}, we find that $\tau$ monotonically increases with increasing $\tilde{\kappa}$, quantifying the slower polar ordering exhibited by more rigid filaments. The saturation value $r$ varies between simulation runs, and its ensemble average increases slightly with increasing $\tilde \kappa$, indicating better polar ordering in more rigid filaments.

We next investigate the effect of changing the area fraction $\phi$ of filaments on the LRO state (Appendix \ref{Appendix B} Fig.~\ref{fig:area_frac}, \ref{fig:area_frac_fit}, \ref{P1_P7}). We see that the lowest area fraction tested ($\phi = 0.1$) takes the longest to reach the long-range orientational order (Fig.~\ref{fig:area_frac_fit}\textbf{A}). When we simulate the systems for much longer times compared to $t_*$, and fit the \revision{ensemble-averaged} polar order parameter to a saturating exponential as we did in Fig.~\ref{fig:steady_state_plots}\textbf{C}, we find that the time scale $\tau$ required to reach a LRO polar steady state monotonically decreases with increasing $\phi$ (Fig.~\ref{fig:area_frac_fit}\textbf{C}). This effect can be attributed to an increase in the rate of filament interactions for higher $\phi$\revision{-}values, giving the system more opportunities for reorientation and hence reaching the steady state sooner. 

\revision{The wide range of plateau $P$-values observed at the time $t_*$ of nematic ordering poses a challenge for isolating the effect of $\phi$ on the timescale for polar ordering. To control for different histories of partial polar ordering, we conducted a set of simulations in which the filaments are initialized in one of two opposing directions, so that the initial global nematic order is perfect, $S_0=1$, and the initial global polar order $P_0$ is precisely specified (Fig.~\ref{P1_P7}). 
A polar order saturation time $T_{P_c}$ was defined as $T_{P_c} = \tilde t_{P_c} + \tau_{P_c}$, where $P_c$ is a chosen threshold $P$-value, $\tilde t_{P_c}$ is the time at which the increasing $P(\tilde t)$ reaches $P_c$, and $\tau_{P_c}$ is the time constant from an exponential fit to $P(\tilde t)$ for all later times. 
By calculating $T_{P_c}$ for an ensemble of simulations over a range of $P_0$ and $\phi$ values, we find that systems with higher area fraction take less time to reach LRO polar order for each $P_0$ (Fig.~\ref{T_all}.) 
Furthermore, the exponential saturation time constant $\tau_{P_c}$ is remarkably consistent across a wide range of $P_0$ values (excluding cases of $P_0 > P_c$) at each value of $\phi$ (Fig.~\ref{tau_all}). This latter result indicates that the late-time polar ordering dynamics have no dependence on the history of nematic ordering other than through a $\tilde t$-shift that depends on the plateau value $P(\tilde t_*)$. These findings are consistent over a range of choices of cutoff value $P_c$ for the exponential fits.
}

Changing the system size $L_x$ does not seem to have any significant effect on the typical time scales needed to reach long-range orientational order ($t_*$) or to reach the LRO active polar steady state ($\tau$) (Appendix \ref{Appendix D} Fig.~\ref{lx}). The plateau values of the polar order seen in Fig.~\ref{fig:steady_state_plots}\textbf{B} are also seen not to depend significantly on $\phi$ and $L_x$. We additionally note that incorporating anisotropic friction into the Brownian dynamics model \cite{Betterton} does not qualitatively change our results, as seen in Appendix \ref{Appendix C}  Fig.~\ref{fig:aniso}.

\section{Discussion and Conclusions}

In this study, our objective was to investigate ordered states in the collective motions of self-propelled, semiflexible filaments.  With a microscopic model lacking attractive or aligning mechanisms between filament pairs, we find that crossovers are essential to the emergence of LRO states. In contrast,  strictly two-dimensional confinement permits only locally ordered polar flocks \revision{\cite{Tanida, Schaller2010, Weitz2015, abkener2013, Betterton, Gromann2020}}. This finding is consistent with a recent experimental study of \textit{E.~coli} bacterial colonies on a surface, which are seen to form LRO nematic states when crossovers are allowed, and polar flocks when crossovers are suppressed \cite{Ghosh2022}. 

When we allow filaments to interpenetrate in order to model crossovers, the symmetry of the LRO state that initially forms is nematic for more rigid filaments and polar for more flexible ones. However, we find that the LRO nematic state, despite its experimental importance, is only a transient state in our simulations. All LRO states eventually evolve into LRO polar states, which we find to be the unique dynamical steady state. The typical time required for emergence of the polar steady state increases monotonically with filament rigidity.

Our results suggest that experimental realizations of LRO active nematic states in microtubule gliding assays are long-lived transient states rather than dynamical steady states. Due to factors such as ATP consumption, gliding assay experiments necessarily have finite lifetimes. For relatively rigid filaments such as microtubules, our model predicts that the timescale $\tau$ for emergence of polar order (and thus, the persistence of an apparently nematic state) may be much longer than the time \revision{$\tilde t$} $\sim t_*$ at which nematic order emerges from an isotropic initial condition. Comparing the typical length and active velocities of microtubules in gliding assay experiments \cite{Memarian} to our simulation, we estimate that microtubule bending rigidity corresponds to $\tilde\kappa \approx 250$ (Appendix~\ref{Appendix E}), which is certainly in the high-$\tilde \kappa$ regime according to our results.

Our results also suggest that in the limit of $\tilde{\kappa} \rightarrow \infty$ (active rigid rods) the timescale for polar ordering may diverge, making the active nematic state dynamically stable \cite{Baskaran2008, Ginelli2010, Wensink2012, abkener2013}. Long-lived, long-range nematic order in \textit{E.~coli} colonies might be interpreted as an effectively stable state in a system of effectively rigid rods \cite{Ghosh2022}.

It remains an open question how strongly our findings depend on the use of periodic boundary conditions and reduced system size compared to typical experiments. Surprisingly, however, we find that changing the system size does not have any significant effect on the typical timescale for emergence of the polar steady state. This suggests that our simulations may be capturing the true physics of the bulk system. We also find that our choice of isotropic friction, at the level of individual beads, is not important to our findings, as introducing anisotropic friction does not qualitatively change our finding that the LRO polar order is the steady state (Appendix~\ref{Appendix C} Fig.~\ref{fig:aniso}). 

A number of interesting questions are raised by the plateaus in polar order \revision{$P(\tilde t)$} that characterize the transient LRO nematic state in systems of more rigid filaments, before the late-time emergence of the LRO polar steady state (Fig.~\ref{fig:steady_state}B and \ref{fig:steady_state_plots}B). These plateaus are clear in \revision{$P(\tilde t)$} plots from individual simulation runs (Fig.~\ref{fig:steady_state}B)\revision{,} even though the ensemble-averaged \revision{$\langle P(\tilde t)\rangle$} appears to be monotonically increasing. We have confirmed that the $P$-value of this plateau has no significant dependence on the system size or on the area fraction of filaments (Fig.~\ref{fig:area_frac_fit}, \ref{lx}). The reason for this plateau in the polar order parameter’s evolution, and the dependence of the plateau value on $\tilde{\kappa}$, remain open questions.

The variability that we observe in plateau $P$-values suggests that transient states may have wide variability in their orientational order, in an ensemble of active systems with identical parameters. As nematic order $S$ saturates at earlier times, the polar order $P$ that happens to develop simultaneously is apparently remembered over the long lifetime of the transient LRO nematic state. Increasing the strength of noise in the simulations may help to erase memory of the details of the nematic ordering process and thus reduce differences in order between transient states of different runs, resulting in a narrower distribution of $P$ values.


In this work, we have only focused on the long-range orientationally ordered states with no density inhomogeneities and established that the LRO nematic state is only a transient state. But gliding assays also spontaneously form a variety of ordered states with non-uniform densities \cite{Memarian, Sumino2012, Huber, Tanida, Betterton}. Our results highlight the potential importance of long-lived transients in active states generally, so it is of interest to investigate whether distinct active behaviors may be found in the time-evolution of other, inhomogeneous active states. 

A full understanding of LRO active states, including the role of filament bending rigidity, requires detailed study of binary collisions \cite{Memarian, Tanida, Huber}, which is the subject of ongoing work. This understanding will complement our study in this work of the effect of area fraction $\phi$. We find that increasing $\phi$ reduces the time required for emergence of the LRO polar steady state. This effect can be attributed to an increase in the rate of collisions with increasing $\phi$, giving the system more opportunities to reorient filaments. Other realistic complications of interest, including polydispersity, chirality, and growth, make the active states of self-propelled, semiflexible filaments a subject of rich phenomenology even in the simplest ordered phases. 

\begin{acknowledgements}
    This material is based upon work supported by the National Science Foundation under Grant No.\ DMR-2225543 and by the Hellman Fellows Fund. A portion of this work was carried out at the Advanced Research Computing at Hopkins (ARCH) core facility (rockfish.jhu.edu), which is supported by the National Science Foundation (NSF) grant number OAC 1920103. We are grateful to Kinjal Dasbiswas, Ajay Gopinathan,  Linda Hirst,  Fereshteh Memarian,  Patrick Noerr, Nathan Prouse, Niranjan Sarpangala, and Fabian Jan Schwarzendahl for helpful discussions. 
\end{acknowledgements}

\begin{appendices}
\section{\label{Appendix A}Model details}
\begin{figure}[h]
    \centering
    \includegraphics[width=8cm]{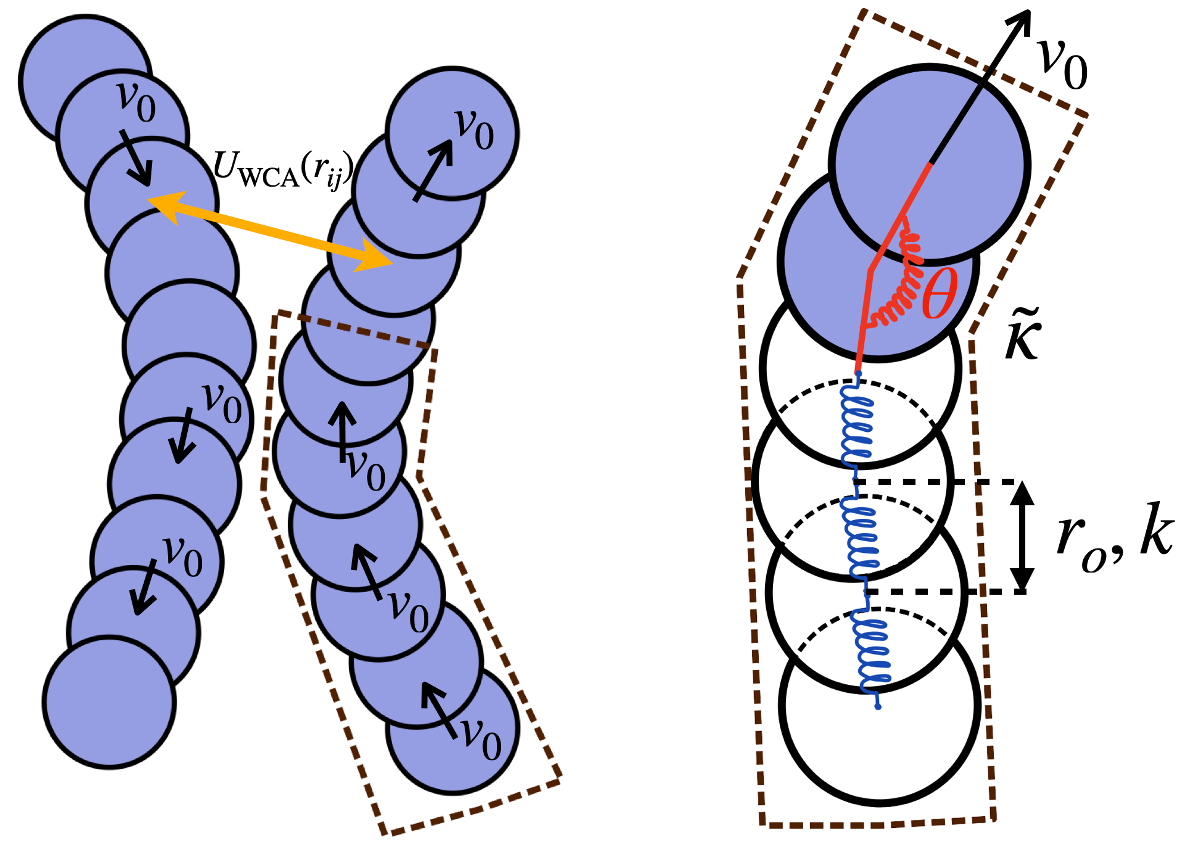}
    \caption{Schematic illustration of forces in the active bead-spring chain model. (Left) Illustration of active self-propulsion and WCA repulsion between pairs of chains. (Right) A zoomed-in view of the region enclosed by the dashed lines at left, to illustrate the stretching and bending spring forces. }
    \label{fig:Picture2}
\end{figure}

Each semiflexible filament is modeled as a bead-spring chain composed of circular disks (beads) connected by Hookean springs with spring stiffness $k = 100$. If $\mathbf{r} = r \hat r $ is the separation vector between the centers of any two connected beads and $r_0$ is the equilibrium spring distance, then the spring force is given by:
\begin{equation}
    \textbf{F}_{\mathrm{stretch}} = -k \left(  r -  r_0 \right) \hat r 
\end{equation}
A discrete approximation to the bending energy for a semiflexible chain is $\frac{1}{2}\kappa\frac{\Sigma_i(\Delta \theta_i)^2}{\Delta s}$ where $\Delta \theta_i = \theta_i - \theta_{i-1}$ is the angle change between points separated by a small distance $\Delta s$ along the contour and $\kappa = \frac{1}{2}l_pk_BT$ is the bending stiffness where $l_p$ is the persistence length of the chain \cite{semiflexible-bending}. We incorporate this bending energy as a potential acting between every set of three consecutive beads ($\Delta s \approx 2r_0$) that penalizes the bending of the chain.
\begin{equation}
    {U}_{\mathrm{bend}} = \frac{1}{2}\tilde{\kappa}\left(\pi - \theta\right)^2
\end{equation}
$\tilde{\kappa} = \frac{\kappa}{\Delta s}$ has the units of energy and will be referred to as the bending stiffness hereafter and $\theta$ is the bond angle between two adjacent linear springs (Fig.~\ref{fig:Picture2}).

The total force excluding viscous drag is given by
\begin{equation}
    \textbf{F}_{\mathrm{total}} = \textbf{F}_{\mathrm{stretch}} - \nabla_\theta U_{\mathrm{bend}} + \textbf{F}_{\mathrm{active}} - \nabla U_{\mathrm{WCA}}
\end{equation}
where $U_{\mathrm{WCA}}$ is given by Eq.~\ref{wca_eq}. The  Brownian (over-damped) dynamics  \cite{Betterton}  updates bead positions through
\begin{equation}
  \textbf{x}(t + dt) = \textbf{x}(t) + \textbf{v}dt +  \sqrt{2Ddt} \boldsymbol{\xi}(t)\revision{,}
  \label{eq.position}
\end{equation}
\begin{equation}
    \textbf{v} =   \textbf{F}_{\mathrm{total}}/\gamma \revision{.}
    \label{eq.fric}
\end{equation}
Here $\textbf{v}$ is the velocity of the bead, $\gamma$ is the drag coefficient, and $\boldsymbol{\xi}(t)$ is the stochastic noise term representing random fluctuations in bead positions, drawn from a zero-mean Gaussian white noise distribution.

\section{\label{Appendix B}Effect of area fraction}
\begin{figure*}
    \centering
    \includegraphics[width=16cm]{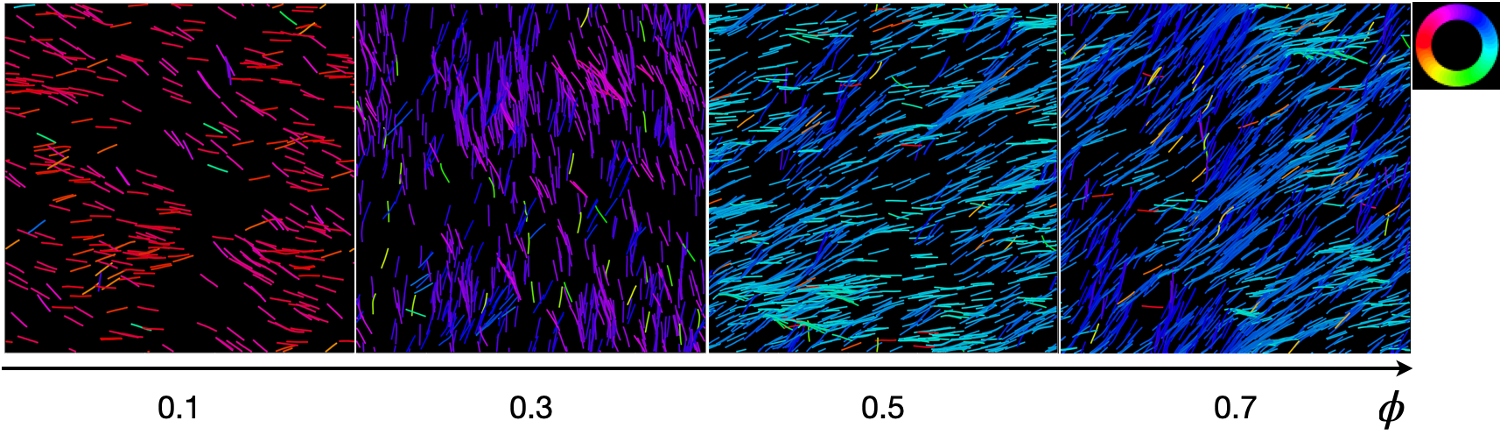}
    \caption{Changing area fraction $\phi$ of filaments. All the snapshots are taken at a time when the system has reached orientational order, which is polar for all cases shown here. 
    The filaments have $\tilde{\kappa} = 10$ and the system size $L_x = 250$.}
    \label{fig:area_frac}
\end{figure*}

We investigated the effect of changing the area fraction of filaments on \revision{the} LRO \revision{order}. 
Fig.~\ref{fig:area_frac} shows snapshots of simulations with different area fractions after they have reached orientational order. All the simulations shown here have filaments with $\tilde{\kappa} = 10$ and $U_{\mathrm{WCA}}$ parameters $\epsilon = 10^{-8}$ and $r_{\mathrm{shift}} = 0.1$ (Fig.~\ref{fig:wca_fig}\textbf{C}). From these snapshots, we see that the symmetry of the LRO state reached for all the area fractions is polar which is consistent with the polar order seen for $\tilde{\kappa} = 10$ in (Fig.~\ref{fig:polar_nematic}) where $\phi = 0.3$. So increasing the area fraction to $\phi = 0.5$ \revision{or} $\phi = 0.7$ does not change the type of orientational order \revision{initially} reached\revision{.}\\  

\begin{figure*}
    \centering
    \includegraphics[width=17.5cm]{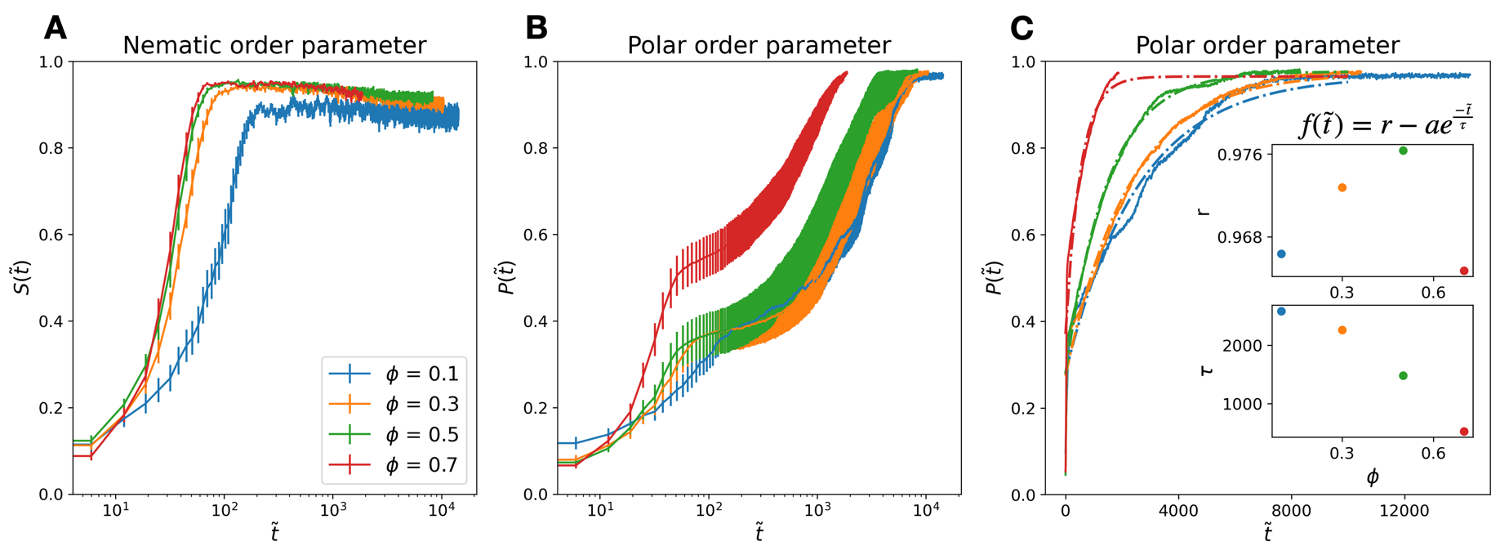}
    \caption{Dependence of orientational order time-evolution upon area fraction $\phi$. Orientational order parameters are averaged over 20-run ensembles with $\tilde{\kappa} = 30$ and with various  $\phi$ values. \textbf{A.} Nematic order time-evolution. \textbf{B.} Polar order time-evolution, with logarithmic scale on the time axis. \textbf{C.} Polar order time-evolution (solid curves), with linear scale on the time axis.  Dashed curves are fits of \revision{$f(\tilde t) = r-ae^{{-\tilde t}/{\tau}}$} to the data. Insets show that $r$ does not depend significantly on $\phi$ (upper inset), while $\tau$ decreases with increasing $\phi$ (lower inset). Time \revision{$\tilde t$} is plotted in units of $\ell/v_0$.}
    \label{fig:area_frac_fit}
\end{figure*}

To understand the steady state behavior of the system, we \revision{examined} 
the order of the system after a long time compared to $t_*$. The global nematic order parameter in Fig.~\ref{fig:area_frac_fit}\textbf{A} reveals that all the simulations reach an orientational order but the system with the lowest area fraction ($\phi = 0.1$) takes the longest time to reach the global orientational order. This is presumably due to filament reorientation interactions being less frequent in the lower-density system. The LRO polar state is still the steady state of the system as seen from Fig.\ref{fig:area_frac_fit}\textbf{B}. The system takes \revision{less} 
time to get to the LRO polar state when the area fraction is higher. Fitting a saturating exponential function \revision{$f(\tilde t) = r-ae^{{-\tilde t}/{\tau}}$} to the mean polar order parameter, we see that $\tau$ decreases with increasing area fraction (Fig.\ref{fig:area_frac_fit}\textbf{C} lower inset) whereas saturation parameter $r$ is not seen to follow any trend with change in area fraction (Fig.\ref{fig:area_frac_fit}\textbf{C} upper inset).

\begin{figure*}
    \centering
    \includegraphics[width=17.5cm]{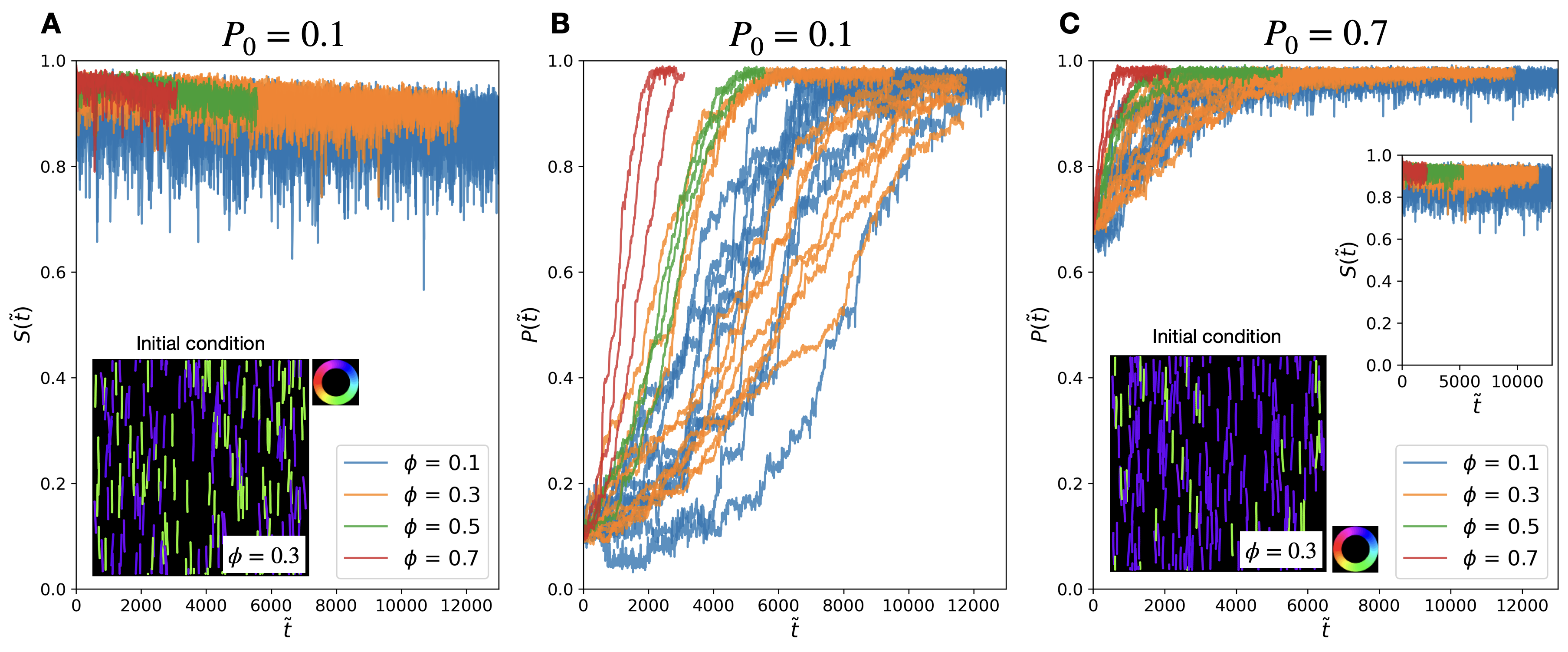}
    \caption{\revision{Time evolution of a system of $\tilde \kappa = 30$ filaments initialized with perfect nematic  order and with specified values of initial polar order $P_0$, for different area fractions $\phi$. \textbf{A.} Nematic order time-evolution for $P_0 = 0.1$. Inset shows a snapshot of the initial condition with filaments oriented either up (purple) or down (green) for an area fraction of $\phi = 0.3$. \textbf{B.} Polar order time-evolution for $P_0 = 0.1$. \textbf{C.} Polar order time-evolution for $P_0 = 0.7$. Inset shows the nematic time-evolution and a snapshot of an initial condition with ($\phi = 0.3$). The ensemble size is $n=10$ for $\phi = 0.1, 0.3$ and $n=3$ $\phi = 0.5, 0.7$, for each value of $P_0$.}}
    \label{P1_P7}
\end{figure*}

\begin{figure*}
    \centering
    \includegraphics[width=17.5cm]{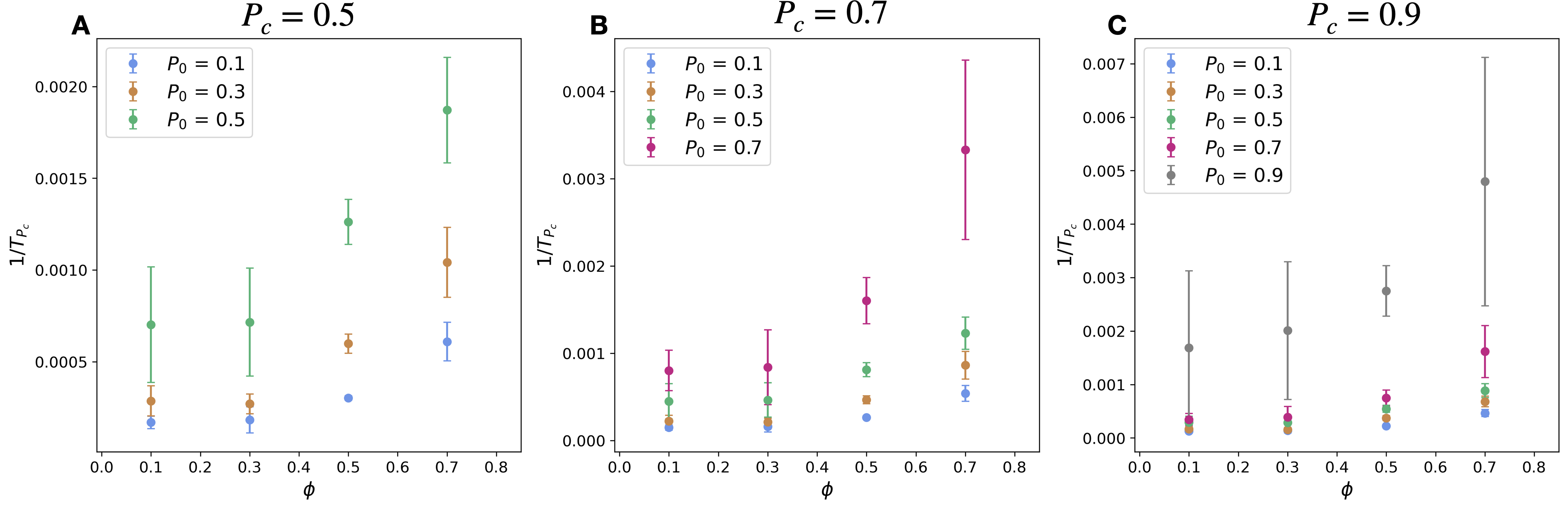}
    \caption{\revision{$1/T_{P_c}$ where $T_{P_c} = \tilde{t}_{P_c} + \tau_{P_c}$ is the total saturation time for different area fraction $\phi$. When $P_0 = P_c$,  $T_{P_c} = \tau_{P_c}$. \textbf{A.} $P_c = 0.5$. \textbf{B.} $P_c = 0.7$. \textbf{C.} $P_c = 0.9$. For all the cutoff $P_c$ values, it can be seen that a higher area fraction leads to lower saturation time for each initial polar order $P_0$. The ensemble size is $n=10$ for $\phi = 0.1, 0.3$ and $n=3$ $\phi = 0.5, 0.7$, for each value of $P_0$.}}
    \label{T_all}
\end{figure*}

\revision{To better understand the dependence of saturation time (from LRO nematic to LRO polar) on $\phi$ separately from its dependence on the plateau value of $P(\tilde t)$, we set up systems with varying area fraction $\phi$ having all filaments oriented initially along one of two opposing directions. These systems have perfect initial nematic order $S_0 = 1$ and controlled polar order $P_0$.  Fig.~\ref{P1_P7} shows the time evolution of nematic and polar order in these systems for two example values of initial polar order, 
$P_0 = 0.1$ and $P_0 = 0.7$. 
Unlike the ensemble-averaged $\left<P(\tilde t)\right>$ examined in Fig.~\ref{fig:steady_state_plots}, $P(\tilde t)$ data for individual runs in these controlled simulations are not generally well-fit by 
a saturating exponential function over the entire simulation time, 
especially for low $P_0$ values such as $P_0 = 0.1$. 
However, because the late-time behavior \textit{is} well-described by a saturating exponential, we can still find a characteristic time scale by defining the total time $T_{P_c}$ to saturation of polar order as $T_{P_c} = \tilde{t}_{P_c} + \tau_{P_c}$ where $\tilde{t}_{P_c}$ is the time taken by the rising polar order $P(\tilde t)$ of each simulation to reach a specific cutoff value $P_c$. For time $\tilde t > \tilde t_{P_c}$ we fit $P(\tilde t)$ to a saturating exponential ($f(\tilde t) = r-ae^{{-\tilde t}/{\tau_P}}$) to obtain the time constant $\tau_{P_c}$. In Fig.~\ref{T_all}\textbf{A}, \textbf{B} and \textbf{C} the cutoff $P_c = 0.5$, $P_c = 0.7$ and $P_c = 0.9$ respectively. Over a wide range of choices of cutoff $P_c$, 
Fig.~\ref{T_all} shows that a higher area fraction leads to lower saturation time for each initial polar order $P_0$. Simulation runs initialized with $P_0 > P_c$ are excluded from this analysis for each choice of $P_c$.}


\begin{figure*}
    
    \centering
    \includegraphics[width=17.5cm]{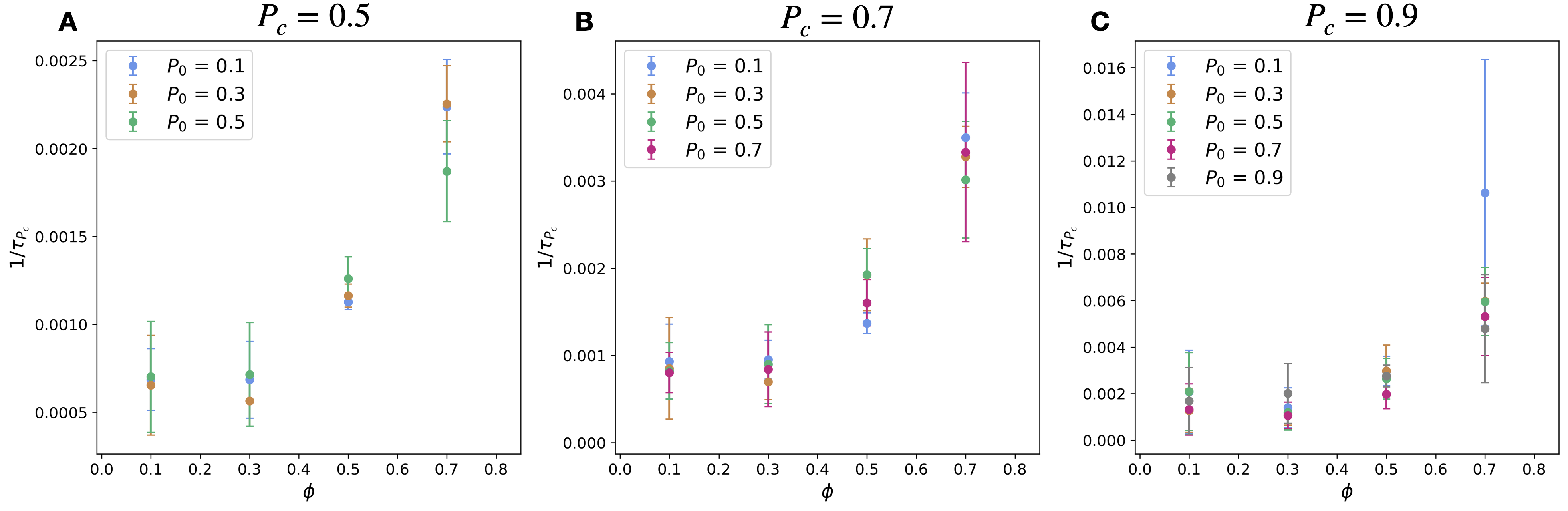}
    \caption{\revision{$1/\tau_{P_c}$, obtained from fitting a saturating exponential $f(\tilde t) = r-ae^{{-\tilde t}/{\tau_P}}$ to $P(\tilde{t})$, is plotted for different initial polar order parameter values $P_0$, for varying area fractions $\phi$. Fits are over the time intervals in which $P(\tilde t) \geq P_c$, with \textbf{A.} $P_c = 0.5$, \textbf{B.} $P_c = 0.7$,  \textbf{C.} $P_c = 0.9$. 
    }}
    \label{tau_all}
\end{figure*}

\revision{Fig.~\ref{tau_all}\textbf{A}, \textbf{B} and \textbf{C} show the fit parameter $1/\tau_{P_c}$ for various values of $\phi$ with cutoffs at $P_c= 0.5, 0.7$, and $0.9$, respectively. It can be seen that the system is memoryless in that the saturation time $\tau_{P_c}$ is similar for all $P_0$. For example, $\tau_{0.7}$ is very similar for $P_0 = 0.7$ (fitting over the entire simulation time-range) as for $P_0 = 0.1$ (in the time-range after $P(\tilde t)$ reaches $0.7$). 
}

\section{\label{Appendix C}Anisotropic friction}
The model described previously assumes isotropic friction on the beads. But as these beads form long filaments with a high aspect ratio, we examine here whether our results are significantly changed by friction anisotropy. For this purpose, we replace Eq.~\ref{eq.fric} with an anisotropic generalization for a particular bead $\alpha$,
\begin{equation}
    v_i^\alpha  = M_{ij}^\alpha  \cdot  F_{\mathrm{total},i}^{\alpha}, 
\end{equation}
where $M_{ij}^\alpha$ is the anisotropic mobility tensor of the $\alpha$th bead \cite{Betterton}:
\begin{equation}
    M_{ij}^\alpha=\frac{1}{\gamma_{\parallel}}{t}_{i}^\alpha \otimes {t}_{j}^\alpha +\frac{1}{\gamma_{\perp}}\left(\delta_{ij} -{t}_{i} ^\alpha \otimes {t}_{j}^\alpha \right).
\end{equation}
Here, $\gamma_{\perp}$ and $\gamma_{\|}$ are the perpendicular and parallel drag coefficients. The symbol $\otimes$ represents the outer product, and $\textbf{t}^\alpha$ is the tangent unit vector to bead $\alpha$, found by 
\begin{equation}
    \mathbf{t}^\alpha =\frac{\mathbf{r}^\alpha+\mathbf{r}^{\alpha+1}}{\left|\mathbf{r}^\alpha+\mathbf{r}^{\alpha+1}\right|}
\end{equation}
where $\mathbf{r}^\alpha$ is the separation vector between the centers of $(\alpha-1)$th and $\alpha$th connected beads, and $1<\alpha<N$ if there are $N$ beads in a chain. For a bead that is located at the ends, $\mathbf{t}^{1} = \mathbf{r}^{2}$ and  $\mathbf{t}^{N} = \mathbf{r}^{N}$. The diffusion coefficient is also anisotropic and for a wormlike chain, it is given by $ D_{ij}^\alpha = k_BT M_{ij}^\alpha$ \cite{Betterton}. This can be used in Equation~\ref{eq.position} to compute the positions of beads at the next timestep.\\

\begin{figure*}
    \centering
    \includegraphics[width=15cm]{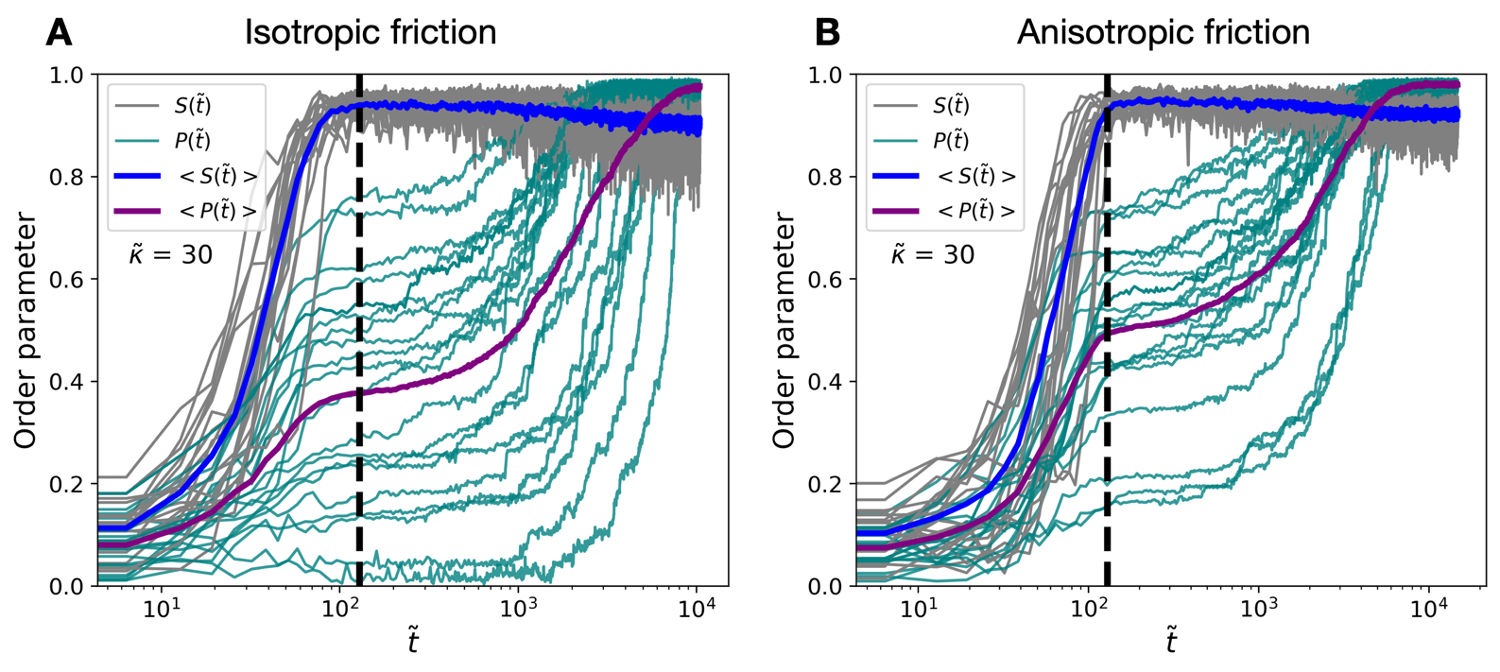}
    \caption{Nematic and polar order parameters are plotted for an ensemble of simulations with $\tilde{\kappa} = 30$, $L_x = 100$, $\phi = 0.3$. \textbf{A.} Isotropic friction. \textbf{B.} Anisotropic friction with an anisotropy $\gamma_{\perp} = 2\gamma_{\parallel}$. Black dashed line represents time $t_*$ as described in the main text.}
    \label{fig:aniso}
\end{figure*}
Using anisotropic friction of $\gamma_{\perp} = 2\gamma_{\parallel}$ in a 20-run ensemble of simulations with $\tilde \kappa = 30$ (Fig.~\ref{fig:aniso}), we see no qualitative difference compared to simulations with isotropic friction as both Fig.~\ref{fig:aniso}\textbf{A} and \textbf{B} reach an LRO polar steady state after being in a transient nematic state at intermediate times. From \revision{$\langle S(\tilde t)\rangle$} values of both the cases, we can see that the case with isotropic friction reaches the LRO nematic state at a slightly earlier time compared to the case with anisotropic friction. However, the case with anisotropic friction is seen to reach the LRO polar steady state sooner. We also find that 
anisotropic friction reduces differences in polar order between transient states of different runs during intermediate times, resulting in a narrower distribution of $P$-values. There is also a slight increase in \revision{$\langle P(\tilde t)\rangle$} at these intermediate times compared to an ensemble with isotropic friction.

\section{\label{Appendix D}Effect of simulation box size}

\begin{figure*}
    \centering
    \includegraphics[width=12.5 cm]{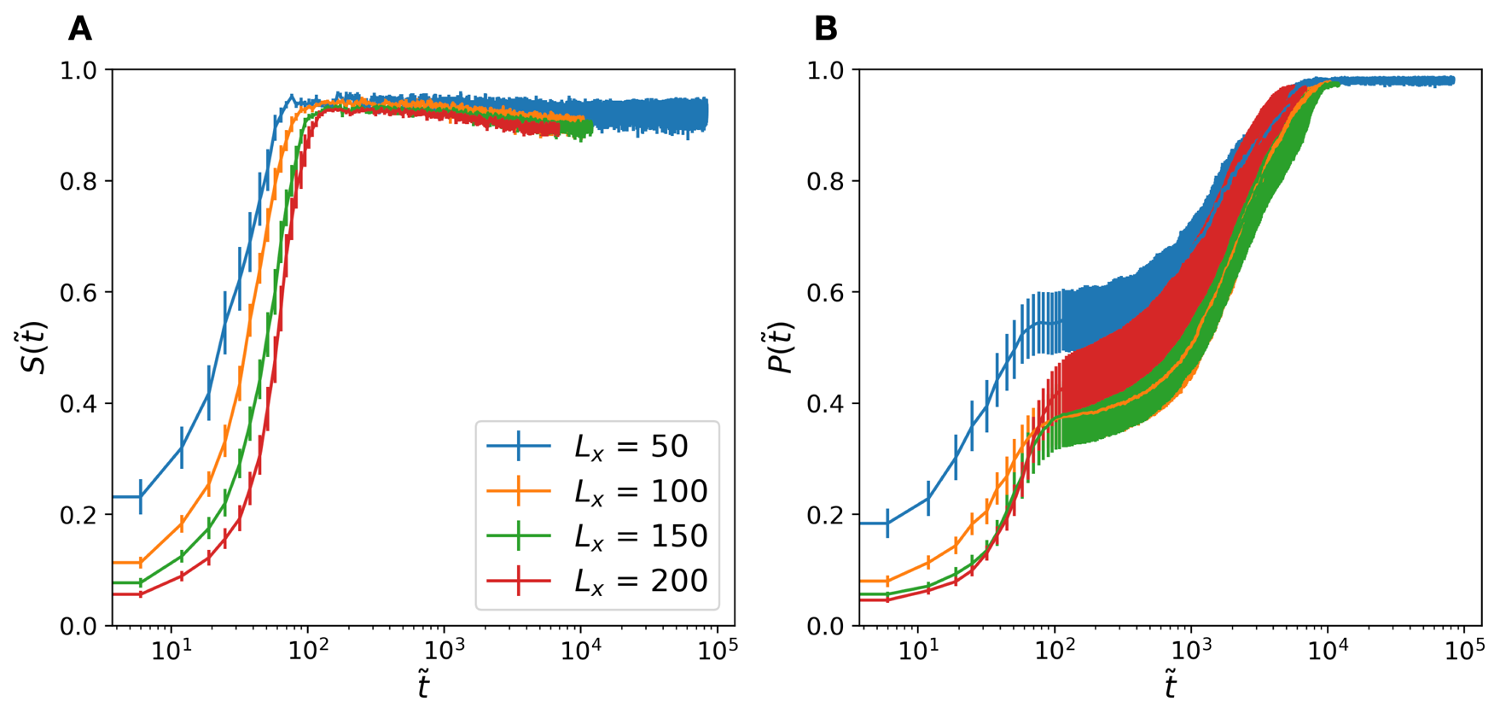}
    \caption{Nematic (left) and polar (right) order parameters are plotted for a 20-run ensemble with $\tilde{\kappa} = 30$, \revision{$\phi = 0.3$}, and four simulation box sizes $L_x$.}
    \label{lx}
\end{figure*}

We examine the effect of changing the simulation box size $L_x$ on the LRO states of the system. We fix $\tilde{\kappa} = 30$ throughout this ensemble of 20 runs for four different $L_x$ values. From Fig.~\ref{lx}\textbf{A}, we see that the initial LRO state is reached at roughly the same time for all system sizes, though there is \revision{a} trend of larger systems taking slightly more time to reach the LRO state. From Fig.~\ref{lx}\textbf{B} we can see that the time to reach the LRO polar steady state does not depend significantly on the system size. Interestingly, we can also see that the plateau value of the polar order in Fig.~\ref{lx}\textbf{B} does not show any monotonic trend with increasing system size. 

\section{\label{Appendix E}Comparison of simulation units to experiments}

\newcommand{\un}[1]{\,\mathrm{#1}}

We provide here a comparison of our simulation units to microtubule (MT) gliding assay experiments on lipid membranes where a depletion agent was not needed to facilitate microtubule bundling \cite{Memarian}. If we equate our simulation filament length $\ell = 15.5 \tilde x$ to the average length of MTs in experiments $ \approx 10 \un{\mu m}$ \cite{Memarian}, we can \revision{obtain} an estimate of \revision{how} the simulation length unit $\tilde x$ \revision{corresponds} to the \revision{experimental} length units: $\tilde x \approx 0.64 \un{\mu m}$. Similarly, the time scaling is set by comparing the active \revision{speed} of the filaments in our simulation ($v_0 = \tilde x / \tilde t = 1$) and the MT velocity in the experiments $\approx 0.5 \un{\mu m/s}$. This implies $\tilde t \approx 1.28s$. To obtain an estimate of mass scaling, we look to the over-damped Brownian dynamics equation \ref{eq.fric}: $[\gamma]= \tilde m / \tilde t$. To 
\revision{estimate} 
the drag coefficient in experiments, we can use the Einstein relation, $D = k_B T/\gamma$ where $D$ is the diffusion constant. From Ref.~\cite{Memarian}, $D \approx 0.1 \un{\mu m^2/s}$ for microtubules in the gliding assay experiment and $k_B T = 4.1 \times 10^{-21}\un{J} $. This gives an estimate of $\gamma = 4.1 \times 10^{-8} \un{kg/s}$ and the mass scaling in our simulations is then $\tilde m = 5.2 \times 10^{-8} \un{kg}$. Thus the simulation units of energy correspond to:
\begin{equation}
    \tilde \epsilon = \frac{\tilde m \tilde x^2}{\tilde t^2} \approx 1.3\times 10^{-20}\un{J}.
\end{equation}
For the bending stiffness $\tilde \kappa$, we obtain a relation to the stiffness of MTs in experiments using the persistence length $l_p \approx 10^{-3} \un{m}$ of MTs. Since bending energy in our model acts between every set of three consecutive beads, we have $\Delta s = 2r_0\tilde x$ (refer to the model details in Appendix~\ref{Appendix A}). Bending stiffness $\tilde \kappa$ is related to energy units $\tilde \epsilon$  by: 
\begin{equation}
    \tilde \kappa = \frac{l_p k_B T}{2 \Delta s \tilde \epsilon} \approx 2.46 \times 10^{2} \un{J }.
\end{equation}
\revision{Thus} the bending stiffness of taxol-stabilized MTs in the gliding assay experiment on a lipid membrane in Ref.~\cite{Memarian} roughly corresponds to a value of $\tilde \kappa \approx 250$ in our simulations. Another common method of stabilizing microtubules is by using guanosine-5$'$ [($\alpha$, $\beta$)-methyleno] triphosphate sodium salt (GMPCPP), and this is known to increase the rigidity of microtubules compared to taxol \cite{Hawkins2013}. The corresponding rigidity in our simulations is  therefore $\tilde \kappa > 250$, and our results for the high-rigidity regime are expected to apply for both MT stabilization methods.
\end{appendices}
\bibliography{bibfile}

\end{document}